\documentclass[aps,prb,twocolumn,longbibliography]{revtex4-2}
\usepackage{epsfig,amsopn}
\usepackage{graphicx}
\usepackage{color}
\usepackage{amsmath,amssymb}
\usepackage{enumerate}
\newcommand\bea{\begin{eqnarray}}
\newcommand\eea{\end{eqnarray}}
\newcommand\beq{\begin{equation}}
\newcommand\eeq{\end{equation}}

\def\nn{\nonumber}

\def\ga{\gamma}

\def\si{\sigma}
\def\Do{\partial}

\def\dg{\dagger}

\def\ra{\rangle}
\def\ua{\uparrow}
\def\da{\downarrow}

\def\th{\theta}

\def\inf{\infty}
\def\sq{\sqrt}

\begin{document}
\title{Field-free transverse response in altermagnet and topological insulator heterojunctions}
\author{Bijay Kumar Sahoo}
\affiliation{School of Physics, University of Hyderabad, Prof. C. R. Rao Road, Gachibowli, Hyderabad-500046, India}
\author{ Abhiram Soori }
\email{abhirams@uohyd.ac.in}
\affiliation{School of Physics, University of Hyderabad, Prof. C. R. Rao Road, Gachibowli, Hyderabad-500046, India}

\begin{abstract}
 We theoretically investigate the transport properties of a topological insulator (TI) connected to two-dimensional altermagnetic (AM) leads in the absence of an external magnetic field. Exploiting the intrinsic time-reversal symmetry breaking of altermagnets, we report the emergence of a field-free transverse differential conductivity in the resulting AM-TI-AM heterostructure. Our analysis reveals that this transverse response is highly sensitive to the orientation of the N\'eel vector: a finite transverse conductivity manifests when the N\'eel vector possesses an out-of-plane component, whereas it strictly vanishes for in-plane orientations ($\theta = \pi/2$) under symmetric chemical potentials ($\mu_t = \mu_b$) due to the restoration of symmetry between the $k_y$ and $-k_y$ momentum channels. Conversely, when asymmetric chemical potentials are applied ($\mu_t \neq \mu_b$), this symmetry is broken, yielding a finite transverse response even for an in-plane N\'eel vector. Furthermore, we demonstrate that while the longitudinal differential conductivity exhibits Fabry-P\'erot-type oscillations as a function of the TI's chemical potential, the transverse conductivity displays a distinct antisymmetric dependence. Notably, the oscillatory behavior of the longitudinal conductivity persists even at finite temperatures. Spatially, the transverse current contribution decays exponentially with position inside the TI region. We also find that both the longitudinal and transverse differential conductivities exhibit periodic oscillations with the junction parameter $\chi_t=\chi_b$. These findings provide a fundamental understanding of field-free transverse responses in spin-momentum locked surface states, highlighting the potential of AM-TI heterojunctions for next-generation spintronic applications.
\end{abstract}

\maketitle

\section{Introduction}
In recent years, a distinct class of quantum materials, known as topological insulators (TIs), has emerged, distinguished by unconventional electronic properties that originate from the topological character of their band structures~\cite{Cheng2011,Kane10}. TIs are novel quantum materials that remain insulating in the bulk yet sustain symmetry-protected boundary states, a consequence of spin–orbit coupling and time-reversal symmetry rather than conventional symmetry breaking~\cite{Cheng2011,Kane10,Sourabh14}. The boundary states typically form Dirac-like bands with spin–momentum locking, so that electron spin is rigidly tied to propagation direction and elastic backscattering from nonmagnetic disorder is strongly suppressed.

In metallic samples, the application of a longitudinal electric bias subject to a magnetic field oriented in the plane can give rise to a voltage along the transverse direction, an effect referred to as the planar Hall effect (PHE). An extensive study has been done on PHE in materials exhibiting spin–orbit coupling subjected to a coplanar magnetic field, and the underlying mechanism is associated with the combined effect of spin-orbit interaction and the Zeeman field~\cite{soori2021,Suri_2021}. In Datta-Das transistors, a transverse current can be generated by supplying spin-polarized electrons to the spin-orbit-coupled central region~\cite{Sahoo2023}, rather than applying a coplanar Zeeman field. 

However, PHE in TIs has also been reported and has emerged as an effective transport probe of their surface states, where spin-momentum locking induces an anisotropic magnetoresponse that produces a transverse voltage subject to a coplanar magnetic field~\cite{Taskin2017,Zheng20,David18}. Unlike the conventional Hall response, this signal arises from symmetry breaking and anisotropic scattering of Dirac surface carriers rather than the Lorentz force, enabling clear identification of surface-dominated transport in topological materials~\cite{Taskin2017,Tewari24,Archit21}. Experimental observations in topological insulator thin films and related theoretical analyses have established the planar Hall response as a distinctive manifestation of the interplay between in-plane magnetic fields and spin-textured surface states~\cite{Zheng20,Tewari24,shafiei25,Rao21}.

A novel class of magnetic materials, referred to as altermagnets (AMs), has been introduced in recent years~\cite{Smejkal22,smejkal22a,smejkal22c}, exhibiting key properties shared by both ferromagnets (FMs) and antiferromagnets (AFMs). AMs host spin-polarized Fermi surfaces, akin to FMs, while maintaining zero net magnetization, as in AFMs. AMs break time-reversal symmetry ($TR$) intrinsically, since the opposite spin sublattices are related through crystal rotation instead of translation. AMs are well-suited for field-free spintronic applications owing to their zero net magnetisation, which suppresses unwanted stray magnetic fields~\cite{Cheng24,Nowak25,sahoo26}.

Motivated by these results, we ask whether a transverse response can arise in AM–TI heterojunctions under field-free conditions. In the present work, we examine transport through an AM–TI–AM junction without any external magnetic field, in which a TI is connected to two-dimensional AMs on both sides. Our results show that a N\'eel vector oriented out of the plane produces a finite transverse differential conductivity, while an in-plane orientation of the N\'eel vector yields no transverse response under symmetricgating of TI surfaces. Furthermore, we find that the transverse differential conductivity is position dependent. Again, we observe that longitudinal differential conductivity shows a Fabry-P\'erot-type oscillation with the chemical potential of TI, while the transverse differential conductivity of TI shows an antisymmetric nature with the same. Moreover, the oscillatory behaviour of the longitudinal conductivity remains robust to finite temperatures. Additionally, we find that the transverse conductivity of the system vanishes at $\th=\pi/2$ when $\mu_t=\mu_b$, due to the restored symmetry between $k_y$ and $-k_y$.

In the theoretical framework of our AM-TI junction, $G_{yx}(x)$ represents a local response function defining the transverse charge current density generated per unit of applied longitudinal bias. Although the model Hamiltonian does not explicitly incorporate transverse voltage probes, a standard experimental realization employing a Hall-bar geometry with open transverse boundaries will experience charge accumulation at its lateral edges driven by this steady-state local transverse current. This accumulation subsequently establishes a position-dependent transverse Hall voltage, directly connecting the calculated $G_{yx}(x)$ to an experimentally measurable local Hall voltage~\cite{soori2021}.  To accurately capture the spatial dependence, the width of the contacts probing the transverse voltage should be small compared to the length scale over which $G_{yx}(x)$ varies with position. 

\section{Details of calculation}

\begin{figure}
    \centering
    \includegraphics[width=9cm]{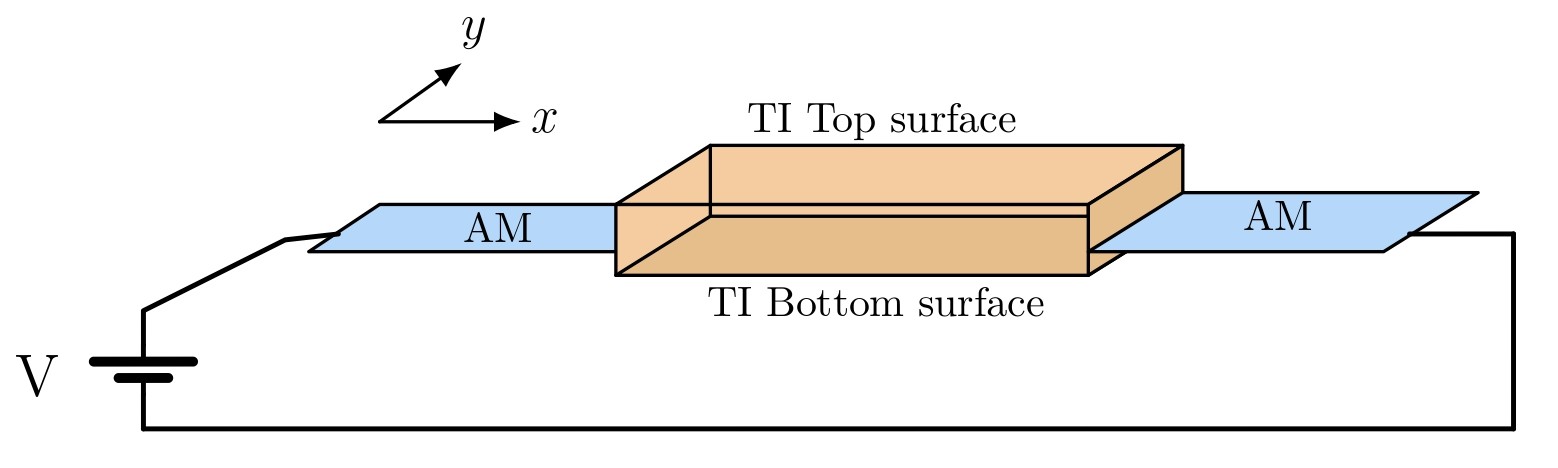}
    \caption{Schematic of the junction: where a TI segment is placed between two AM leads. The system retains translational symmetry in the $y$-direction, whereas both AM leads are semi-infinite in the $x$-direction. The left and right AM leads are subjected to a voltage bias $V$.}
    \label{fig:schem}
\end{figure}

We study a system comprised of AM-TI-AM junctions, where the middle TI has both top and bottom surfaces, as shown in Fig. 1. The system preserves translational symmetry along the $y-$ direction, so the transverse momentum $p_y$ remains conserved and identical across all regions of the system. In this work, we study transport through TI surface states while ignoring the TI side surface states. We consider the TI thickness to be smaller than the electron Fermi wavelength ($\lambda_F$). The AM leads on either sides of TI are semi-infinite, with left AM extends from $x=-\inf$ to $x=0$ and right AM extends from $x=L$ to $x=\inf$.The TI spans the interval $0\le x_p \le L$ and make junctions with left AM along $x=0$, while with right AM along $x=L$. We will use subscripts $t(b)$ for the coordinates of top(bottom) TI surfaces. In reciprocal space, the altermagnetic system is described by the tight-binding Hamiltonian as follows:
\bea
H_A &=& [-2t_0 (\cos k_xa+\cos k_ya)]\si_0 \nn \\
&& +2t_J (\cos k_xa - \cos k_ya)\si_{\th,\phi}
\label{TH-A}
\eea
where, $t_0$ denotes the nearest-neighbor hopping amplitude, $t_J$ characterizes the AM strength, $a$ represents the lattice constant, and $\si_{\th,\phi}=\si_z \cos{\th}+\sin{\th}(\si_x \cos{\phi}+\si_y \sin{\phi})$ with $\si_{x,y,z}$ are Pauli spin matrices. In this work, we exploit the continuum Hamiltonian of AM by expanding the above equation near the band bottom (i.e. $k_x=0$ and $k_y=0$) for $0<t_J<t_0$.

So the Hamiltonian describing our system can be written as
\bea
H &=& [-t_0 a^2(\Do_x^2+\Do_y^2)-\mu_A]\si_0 + t_J a^2(\Do_x^2-\Do_y^2)\si_{\th,\phi},  \nn \\ 
&& {\rm for ~~}x<0, ~~{\rm and }~~x>L, \nn \\
&=& i\hbar v_f (\si_y \Do_{x_t}-\si_x \Do_{y_t})-\mu_t \si_0, ~{\rm for ~~}0<x_t<L,\nn \\
&=&  -i\hbar v_f (\si_y \Do_{x_b}-\si_x \Do_{y_b})-\mu_b \si_0,~  ~{\rm for ~~}0<x_b<L.\nn \\
\eea

Here, $\mu_A$ is the chemical potential in AM, $\mu_{t/b}$ represents the chemical potential for top/bottom TI surfaces, $v_f$ corresponds to the Fermi velocity of electrons in TI.

The dispersion relation for AM and TI surfaces are
\bea
E &=& (t_0-\ga_{\si}t_J) k_x^2 a^2+(t_0+\ga_{\si}t_J) k_y^2 a^2-\mu_A, \\
&& {\rm with~} \ga_{\si}=+,- ~{\rm for~} \si=\ua,\da  {\rm with~respect~to}~~ \si_{\th,\phi}. \nn \\
E &=& -\mu_p \pm \hbar v_f k,~~ {\rm with ~} k=\sqrt{k_x^2+k_y^2},\\
&& ~{\rm and~} p=t,b ~{\rm for~top~and~bottom~TI~surfaces.} \nn 
\eea

The expressions describing charge current densities along the longitudinal and transverse directions in AM and TI regions are as follows:
\bea
J_x^A=\frac{2ea^2}{\hbar} {\rm Im}[\psi_A^\dg (t_0 \si_0-t_J \si_{\th,\phi})\Do_x \psi_A ] \\
J_x^t=-e v_f (\psi_t^\dg \si_y \psi_t)~ {\rm and~} J_x^b=e v_f (\psi_b^\dg \si_y \psi_b) \\
J_y^t=e v_f (\psi_t^\dg \si_x \psi_t)~ {\rm and~} J_y^b=-e v_f (\psi_b^\dg \si_x \psi_b)
\eea

Here, $\psi_A$ denotes the wavefunction within the AM region, while $\psi_{t}$  and $\psi_{b}$ correspond to the TI's top and bottom surface wavefunctions. The charge current is conserved along the longitudinal direction on both sides of the junction at $x=0, L$ (i.e. $J_x^A=J_x^t+J_x^b$). The most general form of boundary conditions that ensures this current continuity can be written as follows: 
\bea
\psi_A=c[M(-\chi_t) \psi_t + M(\chi_b) \psi_b], \nn \\
\frac{2}{\hbar v_f} (t_0 \si_0-t_J \si_{\th,\phi}) \Do_x \psi_A + \chi_A \psi_A =\nn \\
 \frac{i}{c} \si_y. [-M(-\chi_t) \psi_t + M(\chi_b) \psi_b].
\label{BC}
\eea
Here, all the wavefunctions, with their derivatives are evaluated precisely at the interface located at $x=0$. Here, $M(\chi)=e^{i\chi\si_y}$, where the dimensionless quantities $\chi_A,\chi_t$ and $\chi_b$ determine the strength of the delta-potentials positioned infinitesimally near the interface on the AM side, and the corresponding TI's top and bottom surfaces, respectively~\cite{Modak12,Diptiman12,Suri_2021, Soori13}. At $x=L$, the form of the boundary remains the same as in Eq.\ref{BC}, however, the dimensionless quantities $\chi_A,\chi_t$ and $\chi_b$ enter with reversed signs. In this work, we set $c=1$, and all the dimensionless barrier strengths $\chi_t,\chi_b$ and $\chi_A$ are set to zero unless stated otherwise.

 The wavefunction describing an electron with spin $\si$, approaching from the left AM region at an angle $\beta$ relative to $x$-axis with energy $E$ has the form $\psi(x)e^{ik_y y}$ with:
\bea
|\psi_A (x)\ra &=& (e^{i k_{x,\si}x}+r_{\si} e^{-i k_{x,\si}x}) |\si \ra \nn \\&& + r_{\bar {\si} } e^{-i k_{x,\bar {\si}}x} |\bar{\si}  \ra, 
~~ {\rm for ~~}x<0, \nn \\
|\psi_p (x_p)\ra &=& \sum_{p=t,b} [s_{\si,p,+} e^{i k_{x,p,+}x_p} |k_{p,+}\ra \nn \\ && +s_{\si,p,-} e^{i k_{x,p,-}x_p} |k_{p,-}\ra], \nn \\
&& {\rm for ~~}0<x<L, \nn \\
|\psi_A (x)\ra &=& t_{\si} e^{i k_{x,\si}x} |\si \ra + t_{\bar{\si}} e^{i k_{x,\bar {\si}}x} |\bar {\si} \ra, \nn \\
&& {\rm for ~~}x>L,
\eea
where $\si=\ua,\da$ with respect to $\si_{\th,\phi}$, $\bar{\si}$ is the opposite spin to $\si$, ${|\ua\ra}=[\cos{(\th/2)},~ \sin{(\th/2)}e^{i\phi}]^T$, ${|\da\ra}=[-\sin{(\th/2)},~\cos{(\th/2)}e^{i\phi}]^T$, $k_{x,\si}=\sqrt{(E+\mu_A)/(t_0-\ga_{\si}t_J)}\cos{\beta}$ and $k_{y,\si}=\sqrt{(E+\mu_A)/(t_0 +\ga_{\si}t_J)}\sin{\beta}$ with $\ga_{\si}=+,-$ for $\si=\ua,\da$, $k_{x,\bar{\si}}$ is calculated from $\bar{\si}$-spin band dispersion in AM taking same value of $k_y$ as in the incident region, $k_{x,p,\pm}$ are obtained from the dispersion relation of $p$-TI, $|k_{x,p,\pm} \ra$ are the eigen spinors of $p$-TI obtained from TI Hamiltonian for electron with wavenumber $(k_{x,p,\pm}, k_y)$ and the scattering coefficients $r_{\si}$, $r_{\bar{\si}}$, $s_{\si,p,\pm}$, $t_{\si}$, $t_{\bar{\si}}$ are to be determined by using the boundary condition at $x=0,L$.

Under an applied voltage bias, the longitudinal and transverse differential conductivities are defined as $G_{xx} = dI_x/dV$ and $G_{yx}(x) = dI_y/dV$, respectively. Here, $I_x$ and $I_y$ represent the total current densities along the $x$- and $y$-directions, encompassing the contributions from all incident modes within the infinitesimal bias window. The expressions for these conductivities are
\bea
G_{xx} &=&\frac{e}{8\pi^2a^2\sq{t_0^2-t_J^2}} \sum_{\si=\ua,\da}\int_{-\pi/2}^{\pi/2} d\beta~J_{x,\si}^A, \label{Gx}\\
G_{yx}(x) &=&\frac{e}{8\pi^2a^2\sq{t_0^2-t_J^2}} \sum_{\si=\ua,\da}\int_{-\pi/2}^{\pi/2} (J_{y,\si}^t+J_{y,\si}^b)~ d\beta,  \nn \\
&& {\rm for ~~}0<x<L. 
\label{Gy}
\eea

Here, the voltage bias V is implemented by introducing a difference in the electrochemical potentials between the left and right AM leads (note that this is different from the chemical potential $\mu$ that enters the Hamiltonian) in Landauer formalism. The differential conductivities are then obtained at zero temperature by evaluating the current carried by the scattering states within the infinitesimally small energy window opened by $dV$. For the normalization of the scattering states, the incident wavefunctions in the AM leads are normalized to carry a unit probability flux along the longitudinal $x$-direction. The prefactors in Eqs.~\ref{Gx} and \ref{Gy} arise from converting the summation over the transverse momentum $k_y$ into an integral over the incident angle $\beta$, incorporating the density of states of the incident AM lead and the flux normalization of the incident modes.

\section{Results and Analysis}
In this work, we use realistic parameters for numerical calculations motivated by the known properties of $RuF_4$ as an AM candidate~\cite{Milivojević_2024} and $Bi_2Se_3$ as a TI candidate~\cite{Cheng2011}. The parameters are: $t_0=250$meV, $t_J=0.65t_0$, $a=5$\AA, $v_f=3.2t_0a/\hbar$.
We work in the limit $\mu_t=\mu_b$. The electron charge $e = 1.602 \times 10^{-19}$~C and the reduced Planck constant $\hbar = 1.054 \times 10^{-34}$~J$\cdot$s. 
\begin{figure}
    \includegraphics[width=4.2cm]{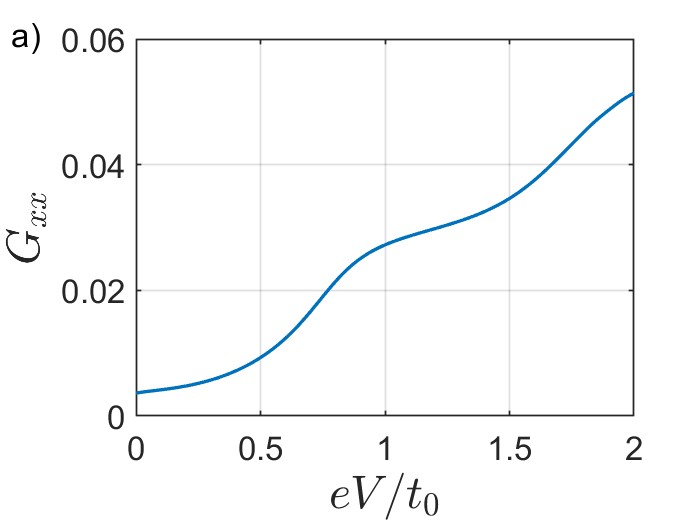}
    \includegraphics[width=4.2cm]{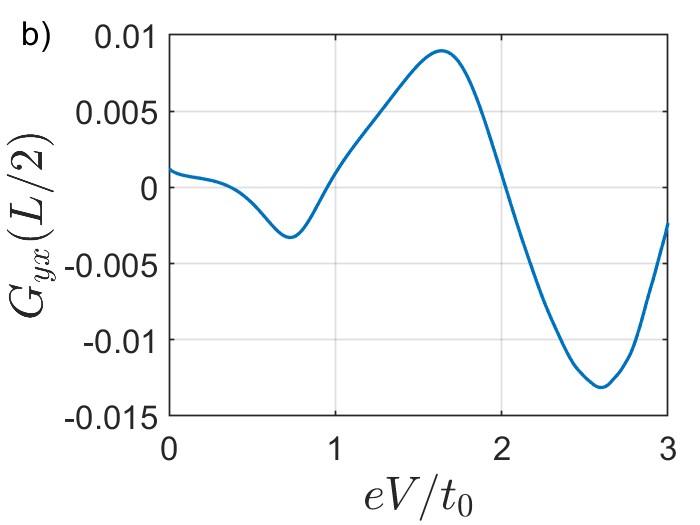}
    \caption{Differential conductivity along longitudinal direction $G_{xx}$ in (a) and along transverse direction $G_{yx}$ at $x=L/2$ in (b) versus bias. The conductivity values presented here are scaled in units of $e^2/{\hbar a}$. Parameters: $t_J=0.65t_0,~v_f=3.2t_0a/\hbar,~\mu_A=0.1t_0,~\mu_t=\mu_b=0.2t_0,~L=10a,~\chi_A=0,~\chi_t=\chi_b=0,~\th=\phi=0.$}
    \label{fig:GvsE}
\end{figure}

We study the bias dependence of longitudinal and transverse differential conductivity ($G_{xx}$ and $G_{yx}$) in Fig~\ref{fig:GvsE}(a) and Fig~\ref{fig:GvsE}(b) for $\mu_A=0.1t_0$, $\mu_t=\mu_b=0.2t_0,~L=10a,~\th=\phi=0$. In Fig~\ref{fig:GvsE}(a), as the bias increases, $G_{xx}$ rises owing to the higher density of states in the TI segment. We observe that $G_{yx}(L/2)$ in Fig.~\ref{fig:GvsE}(b) varies nonmonotonically with the applied bias and exhibits oscillations, which can be attributed to Fabry-P\'erot-like interference effects in the TI region. The dominant contribution to the transverse current arises from propagating modes confined to a narrow window of incidence angles near normal incidence, and this angular window broadens as the bias increases.  
Multiple reflections of the electron wave functions inside the TI segment generate alternating constructive and destructive interference.
This leads to the oscillatory behaviour of $G_{yx}$ as a function of bias. With increasing bias, the minima deepen as the TI Fermi surface increases in size, thereby supporting a greater number of out-of-plane plane-wave states.

\begin{figure}
    \centering
    \includegraphics[width=4.2cm]{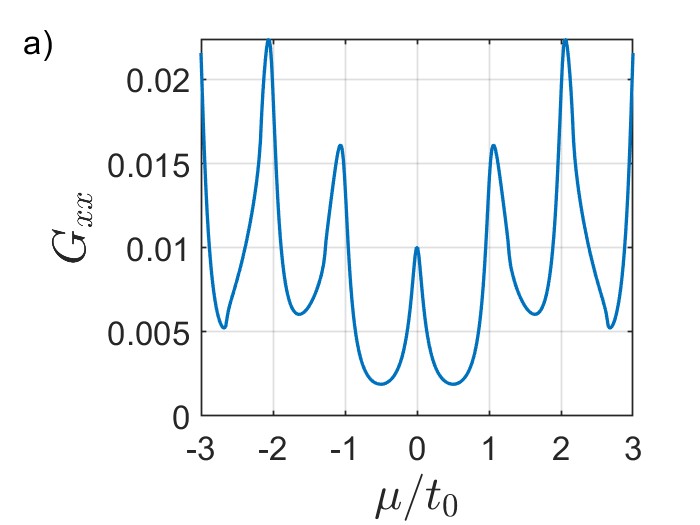}
    \includegraphics[width=4.2cm]{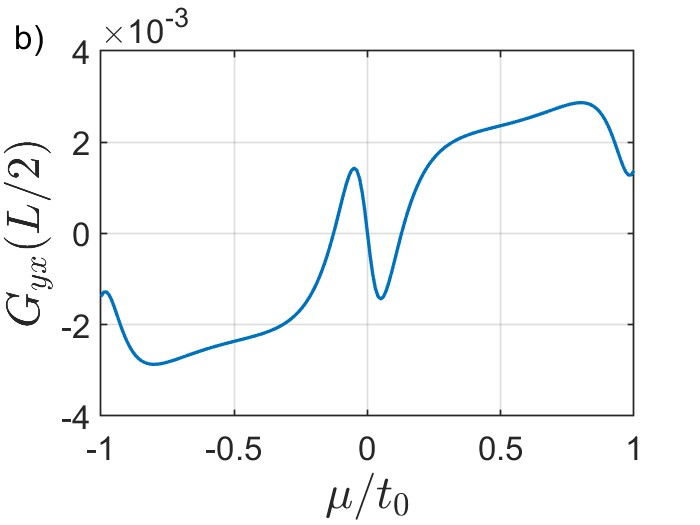}
    \caption{Longitudinal differential conductivity $G_{xx}$ in (a) and transverse differential conductivity $G_{yx}$ at $x=L/2$ in (b) versus the chemical potential of TI for $\mu_t=\mu_b=\mu$ in units of $e^2/{\hbar a}$ at zero bias. The rest of the parameters are kept identical as in Fig~\ref{fig:GvsE}.}
    \label{fig:Gvsmu}
\end{figure}

Now, we study the behaviour of $G_{xx}$ and $G_{yx}$ with chemical potential of TI for $\mu_t=\mu_b=\mu$ at zero bias in Fig.~\ref{fig:Gvsmu}. We observe in Fig.~\ref{fig:Gvsmu}(a) that $G_{xx}$ oscillates with $|\mu|$, with peaks of the oscillation increasing with $\mu$ due to the increase in DOS in TI with $|\mu|$. These oscillatory patterns arise from quantum interference, in which waves reflecting back and forth within the finite TI region interfere either constructively or destructively depending on the phase they acquire, a phenomenon known as Fabry-P\'erot interference (FPI)~\cite{Soori12,Liang2001}. For normally incident electrons, the FPI condition for constructive interference is $\Delta k=\pi/L$, where $\Delta k$ is the separation between two adjacent peaks at $\mu_1$ and $\mu_2$. The value of $\Delta k$ calculated from the above mentioned condition is $0.314/a$, which in close agreement with $0.312/a$ obtained numerically from the results of Fig.~\ref{fig:Gvsmu}(a) by $\Delta k=(\mu_1-\mu_2)/{\hbar v_f}$. The results in Fig.~\ref{fig:Gvsmu} (b) demonstrate that $G_{yx}(L/2)$ exhibits an anti-symmetric behaviour with $\mu$. This anti-symmetric nature can be understood from the TI Hamiltonian. Upon changing the chemical potential as $\mu \rightarrow -\mu$, the surface Hamiltonians associated with the TI's top and bottom surfaces transform into one another up to an overall minus sign, namely  $H_{t/b} \rightarrow -H_{b/t}$. As a consequence, the helicity of the spin-momentum locking on the two surfaces is inverted, which in turn reverses the direction of the transverse current.  Accordingly, $G_{yx}(L/2)$ changes sign under reversal of $\mu$.

\begin{figure}
    \centering
    \includegraphics[width=7cm]{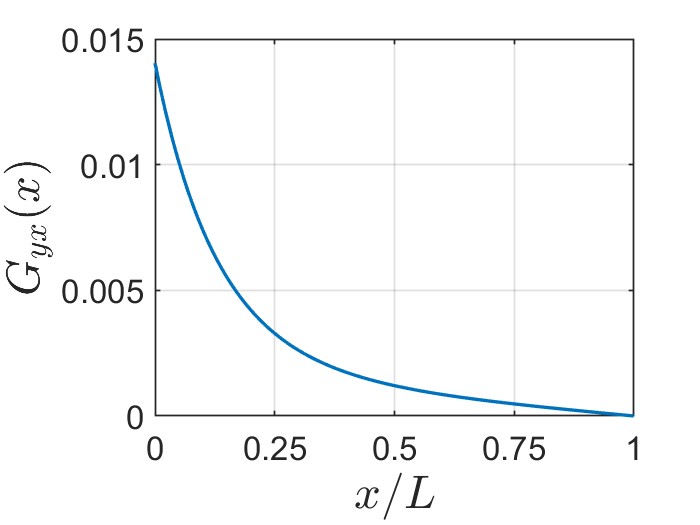}
    \caption{ Transverse differential conductivity $G_{yx}$ of TI as a function of location at zero bias in units of $e^2/{\hbar a}$. Here $\mu_t=\mu_b=0.2t_0$ and all remaining parameters are kept identical as in Fig~\ref{fig:GvsE}. }
    \label{fig:Gvsx}
\end{figure}
In Fig.~\ref{fig:Gvsx} we plot zero-bias transverse differential conductivity $G_{yx}$ of TI versus position for $\mu_t=\mu_b=0.2t_0$, and the rest of the parameters are kept the same as in Fig~\ref{fig:GvsE}. We find that $G_{yx}$ decays exponentially with position, as the wave numbers in the TI for this choice of parameters are complex over a certain range of incidence angles. The contribution to the transverse current is greater in that range of incident angles, leading to exponential decay of the transverse current with position in TI.\\

\begin{figure}
    \centering
    \includegraphics[width=7cm]{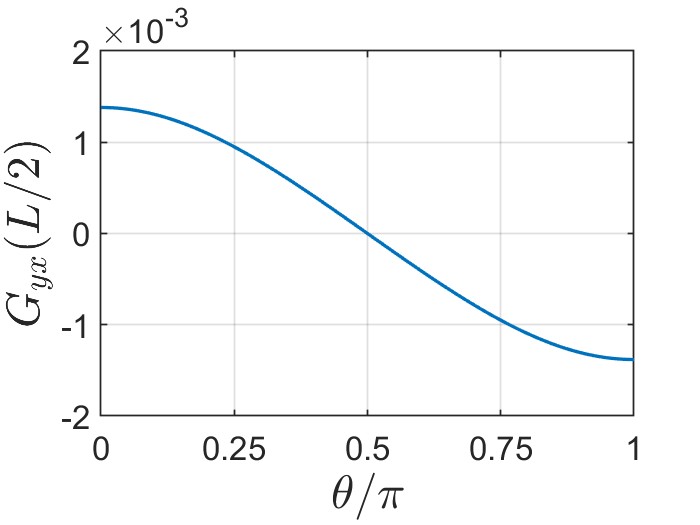}
\caption{Transverse differential conductivity $G_{yx}$ at $x=L/2$ at zero bias  versus $\th$ (angle made by N\`eel vector with z-axis) in units of $e^2/{\hbar a}$ for $\mu_t=\mu_b=t_0$. All remaining parameters are kept identical as in the Fig~\ref{fig:GvsE}. }
    \label{fig:Gvsth}
\end{figure}

We next investigate the dependence of the differential conductivities $G_{xx}$ and $G_{yx}$ on the Néel vector orientation, parametrised by $\th$ and $\phi$, as shown in Fig.~\ref{fig:Gvsth}. We find that both conductances are independent of the azimuthal angle $\phi$, while they exhibit a clear dependence on the polar angle $\th$. This follows from the boundary condition: upon acting with $\si_y$ on Eq.~\ref{BC}, the right-hand side becomes proportional to $\si_0$, while $\si_x-$ and $\si_y-$dependent terms on the left-hand side remain explicitly independent of $\phi$. In the TI surface Hamiltonian, $\si_x$ is coupled to $k_y$, while $\si_y$ coupled to $k_x$. Consequently, neither $G_{xx}$ or $G_{yx}$  exhibits any $\phi$ dependence.

The angular dependence on $\th$ is reflected more prominently in the transverse response. As seen in Fig.~\ref{fig:Gvsth}, the transverse differential conductivity vanishes at $x=L/2$ when $\th=\pi/2$. This can again be understood from Eq.~\ref{BC}. For $\th=\pi/2$, applying $\si_y$ removes the $\si_x$-dependent contribution from the left-hand side, thereby restoring symmetry between the $k_y$ and $-k_y$ channels and thereby making $G_{yx}=0$. In addition, we find from  Fig.~\ref{fig:Gvsth} that $G_{yx}$ reaches its largest magnitude when the direction of N\'eel vector is along $\pm z$, since this configuration produces the strongest inequivalence between the $k_y$ and $-k_y$ sectors.

\begin{figure*}
    \centering
    \includegraphics[width=5.8cm]{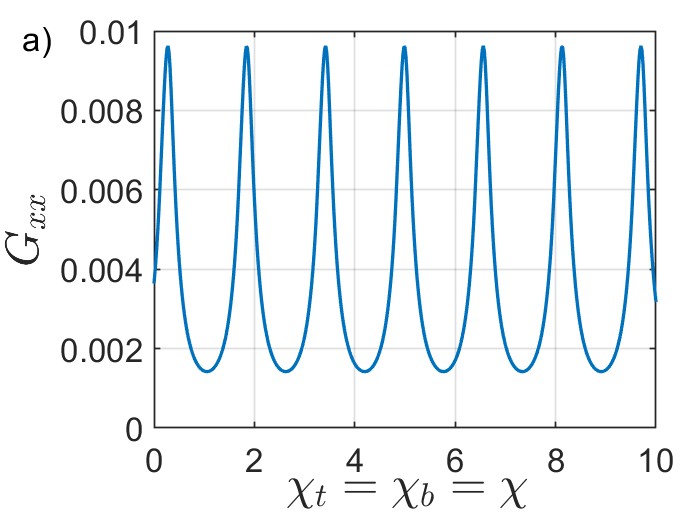}
    \includegraphics[width=5.8cm]{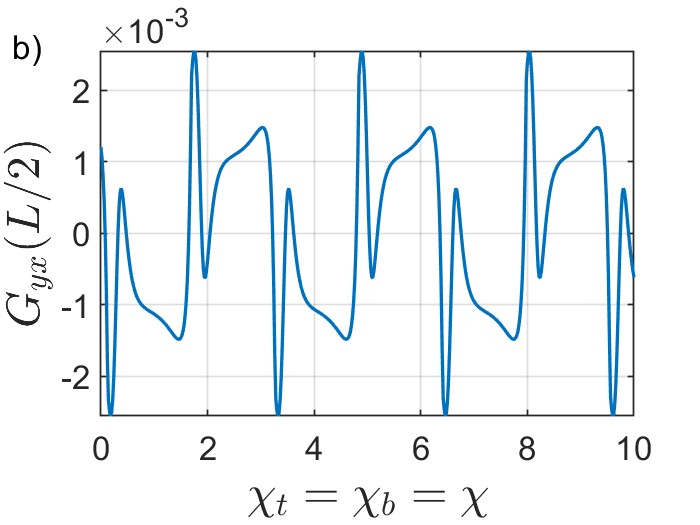}
    \includegraphics[width=5.8cm]{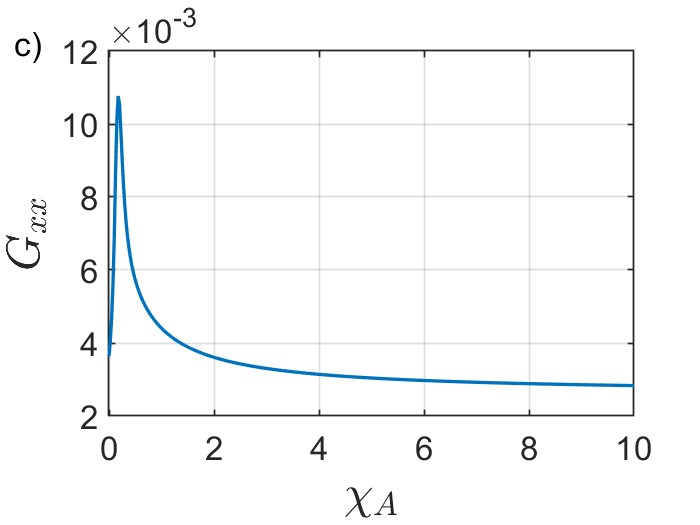}
\caption{Longitudinal differential conductivity $G_{xx}$ in (a) and transverse differential conductivity $G_{yx}$ at $x=L/2$ in (b) versus $\chi_t=\chi_b=\chi$, (c) Longitudinal conductivity versus $\chi_A$. All conductivities are evaluated at zero bias in units of $e^2/{\hbar a}$ for $\mu_t=\mu_b=0.2t_0$. All remaining parameters are kept identical as in the Fig~\ref{fig:GvsE}.}
    \label{fig:Gvschit}
\end{figure*}

Figure~\ref{fig:Gvschit} illustrates the behaviour of differential conductivities as a function of the junction barrier strength for identical top and bottom surface barriers $\chi_t=\chi_b=\chi$ at zero bias. As shown in Fig.~\ref{fig:Gvschit}(a), the longitudinal conductivity $G_{xx}$ oscillates periodically with $\chi$. This behaviour stems from the boundary condition in Eq.~\ref{BC}, where the barrier enters via the phase factor $e^{i\chi \si_y}$~\cite{Diptiman12}. As electrons traverse the junction, they acquire a phase shift, producing conductivity peaks whenever the accumulated phase equals an integer multiple of $2\pi$, yielding an oscillation periodicity of $\approx \pi/2$. The transverse conductivity $G_{yx}(L/2)$, depicted in Fig.~\ref{fig:Gvschit}(b), likewise exhibits periodic oscillations, but with a periodicity $\pi$-  double that of $G_{xx}$. This difference arises because longitudinal electrons encounter the barrier twice (at $x = 0$ and $x = L$), whereas transverse electrons evaluated at $x = L/2$ encounter the barrier only once (at $x = 0$). This effectively doubles the oscillation period of $G_{yx}$ relative to $G_{xx}$ with $\chi$. As shown in Fig.~\ref{fig:Gvschit}(c), the longitudinal conductivity $G_{xx}$ initially increases with increasing barrier strength $\chi_A$, exhibits a peak, and then  saturates at higher values. This persistent saturation occurs because the TI region acts as a cavity for incident electrons, where resonant transmission at specific incidence angles sustains a finite $G_{xx}$ even in the strong-barrier regime (large $\chi_A$).

We also study the effect of temperature on the conductivity. At finite temperature, the expression for conductivity gets modified as follows:
\bea
G_{xx} &=&\frac{e}{8\pi^2a^2\sq{t_0^2-t_J^2}} \sum_{\si=\ua,\da}\int \int_{-\pi/2}^{\pi/2} d\beta~J_{x,\si}^A~f_T(E)~dE, \nn \\
\label{GT}
\eea
Here, $f_T(E)= {\rm sech}^2(\frac{E-eV}{2k_BT})/(4k_BT)$ with $T$ being the temperature and $V$ is the bias voltage.
In Eq.~\ref{GT}, the lower limit of the energy integration is fixed at the band bottom ($E_b = -\mu_a$) of AM owing to the absence of states below this energy, while the upper limit is taken to be on the order of several $k_B T$.

\begin{figure}
    \centering
    \includegraphics[width=7cm]{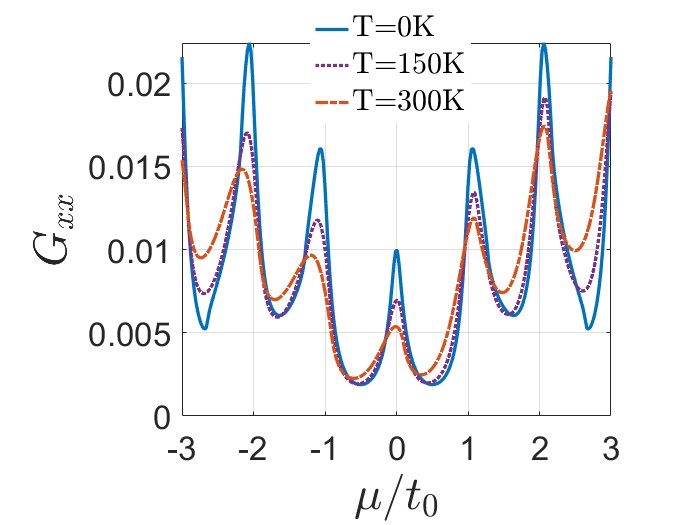}
    \caption{ Zero bias longitudinal differential conductivity $G_{xx}$ versus the chemical potential of TI for $\mu_t=\mu_b=\mu$ for different temperatures($T$) in units of $e^2/{\hbar a}$. All remaining parameters are kept identical as in Fig~\ref{fig:GvsE}. }
    \label{fig:Gvsmu_T}
\end{figure}

In Fig.~\ref{fig:Gvsmu_T}, we plot $G_{xx}$ versus chemical potential of TI for $\mu_t=\mu_b=\mu$ for different temperatures as shown in the legend. We can see the Fabry-P\'erot oscillations of $G_{xx}$ with $\mu$ remain robust against the temperature, though the peaks of the oscillations decrease with temperature.

\begin{figure}
    \centering
    \includegraphics[width=7cm]{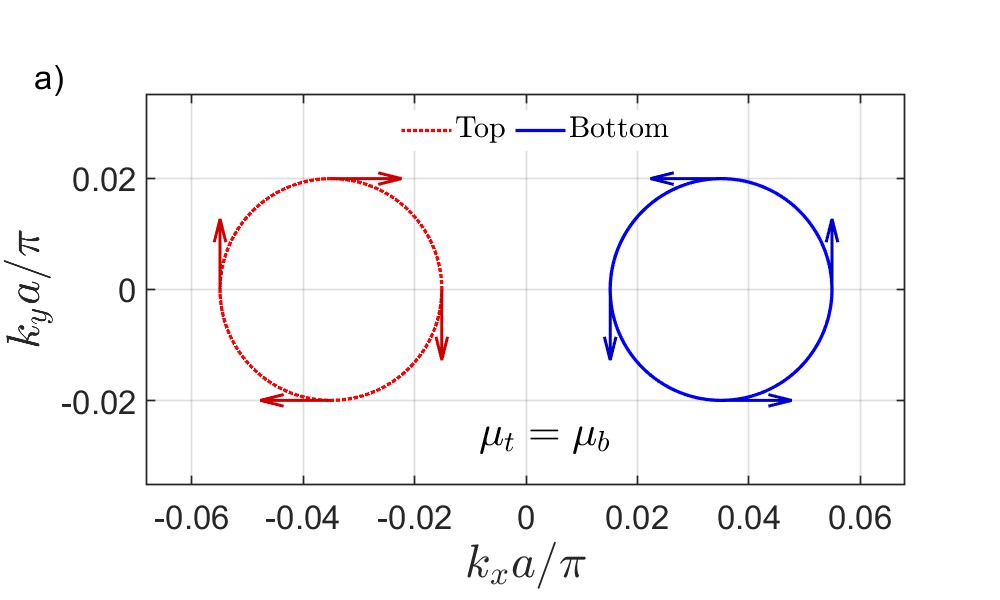}
    \includegraphics[width=7cm]{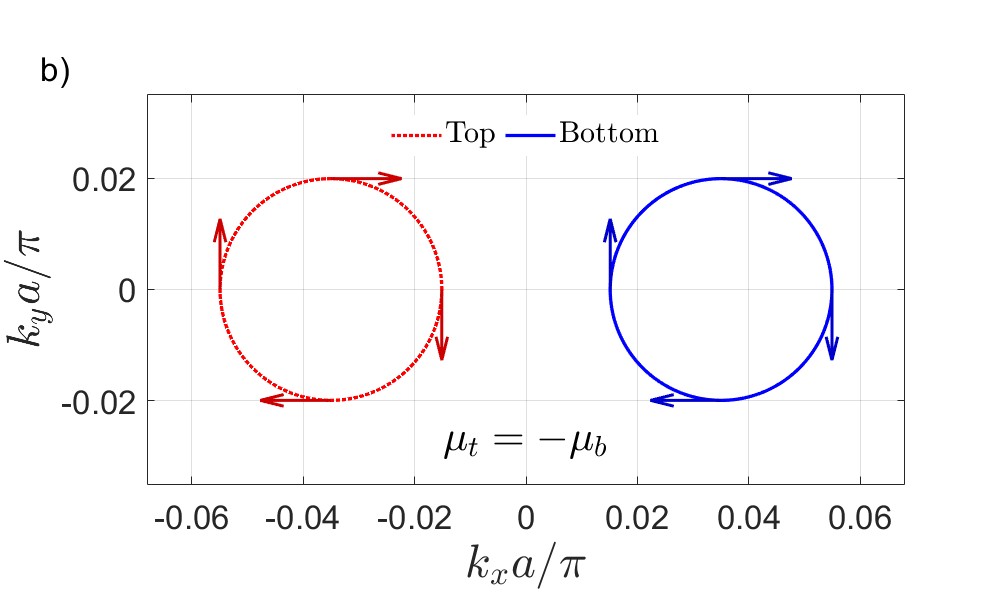}
\caption{ Fermi surface plot with spin momentum locking of top and bottom surface of TI:(a) for $\mu_t=\mu_b=0.2t_0$ and (b) for $\mu_t=-\mu_b=0.2t_0$. Other parameters are same as in Fig~\ref{fig:GvsE}. }
    \label{fig:TI_FS}
\end{figure}

We also performed calculations for asymmetric chemical potentials ($\mu_t \neq \mu_b$) and observed that, for $\theta = \pi/2$, the transverse conductivity $G_{yx}(L/2)$ becomes non-zero, in contrast to the symmetric case ($\mu_t = \mu_b$) where it vanishes. This behaviour can be understood from Fig.~\ref{fig:TI_FS}, which depicts the spin-momentum locking on the top and bottom TI Fermi surfaces (indicated by arrows). For $\mu_t = \mu_b$, the spin-momentum locking on the two surfaces is opposite (see Fig.~\ref{fig:TI_FS}(a)), preserving the combined symmetry of $k_y \to -k_y$ inversion and layer exchange. As a result, the transverse currents along the $y$-direction on the top and bottom surfaces are equal in magnitude but opposite in sign, yielding zero net conductivity. Conversely, for $\mu_t = -\mu_b$, the spin-momentum locking on both surfaces points in the same direction (Fig.~\ref{fig:TI_FS}(b)), breaking $k_y \to -k_y$ symmetry and giving rise to a finite $G_{yx}$.

\section{Discussion}

In this work, the use of AM leads has a distinct practical advantage over conventional ferromagnets (FMs). Conventional FMs possess a net macroscopic magnetisation that inevitably generates stray magnetic fields. When it is interfaced with a TI, these stray fields can inadvertently open a gap in the topological surface states, complicating the field-free transport process.  AMs exhibit strong spin polarisation in momentum space (providing the necessary time-reversal symmetry breaking) with zero net magnetisation in real space. Consequently, they induce the transverse response without any stray fields that would disrupt the TI surface states. Furthermore, the spin splitting in AMs is strongly momentum-dependent. So the spin polarisation of the injected electrons couples to the TI’s spin momentum locking in a highly anisotropic manner that is wholly distinct from the uniform spin-injection of a standard ferromagnet.

Our theoretical formulation treats transport across the top and bottom TI surface states independently, neglecting both inter-surface hybridization and contributions from the side surfaces. This approach is physically justified by assuming the TI lies within an intermediate thickness regime. To safely neglect hybridization, the TI must be thicker than the surface-state penetration depth into the bulk. Concurrently, to suppress side-surface transport, the thickness must remain smaller than the large Fermi wavelength ($\lambda_F$). In this limit, quantum confinement eliminates propagating modes on the side boundaries, rendering their contribution to the total current negligible. This intermediate thickness condition is readily satisfied when the chemical potential is tuned near the Dirac point.

\section{Summary and Conclusion}
In the present work, we have studied  field-free transverse transport in AM–TI–AM heterostructures. Notably, a finite transverse response emerges in the TI without any applied magnetic field, with the altermagnets effectively providing the required symmetry breaking.  We observe that the system exhibits nonzero transverse conductivity when the AM Néel vector lies perpendicular to the plane, whereas a coplanar Néel vector yields a zero transverse response under symmetric gating of the TI surfaces.  We find that longitudinal differential conductivity shows a Fabry-Pe\'rot type oscillations with the chemical potential of TI, while transverse conductivity of TI shows anti-symmetric behaviour with the same. Furthermore, these oscillations in longitudinal conductivity remain pronounced up to room temperature ($T = 300\text{ K}$), provided $k_B T$ remains smaller than the bulk bandgap of TI. This thermal stability underlines the experimental feasibility of observing field-free transverse responses in realistic $\mathrm{AM}\text{--}\mathrm{TI}\text{--}\mathrm{AM}$ heterostructures. Additionally, we find that the transverse conductivity of TI depends on the spatial location and decays exponentially with position. We also analyse its dependence on the N\'eel vector orientation and find that transverse conductivity is independent of azhimuthal angle $\phi$, but varies with polar angle $\th$ for $\mu_t=\mu_b$. This transverse conductivity disappears at $\th=\pi/2$ for $\mu_t=\mu_b$ owing to the restoration of $k_y \to -k_y$ symmetry. In contrast, it remains non-zero at $\theta=\pi/2$ for asymmetric conditions ($\mu_t \neq \mu_b$) due to the breaking of $k_y \to -k_y$ symmetry. We also observe that both the longitudinal and transverse differential conductivities exhibit periodic oscillations with the junction parameter ($\chi_t = \chi_b=\chi$), with the former displaying a periodicity of $\pi/2$ and the latter a periodicity of $\pi$.

 Although reorienting the N\'eel vector is experimentally challenging, recent studies have demonstrated viable approaches for its control in AMs~\cite{han2025,Sayan26}.  These recent advances indicate that the predicted phenomena may be amenable to experimental verification in the near future.

\section{Acknowledgements}
AS and BKS  thank Science and Engineering Research Board, India (now Anusandhan National Research Foundation, India) Core Research grant (CRG/2022/004311) for financial support.  AS thanks University of Hyderabad for funding through Institute of Eminence Professional Development Fund. BKS thanks the Ministry of Social Justice and Empowerment, Government of India, for a fellowship through NFOBC.
\bibliography{ref_TI}

@article{Modak12,
  title = {Spin injection into a metal from a topological insulator},
  author = {Modak, S. and Sengupta, K. and Sen, Diptiman},
  journal = {Phys. Rev. B},
  volume = {86},
  issue = {20},
  pages = {205114},
  numpages = {6},
  year = {2012},
  month = {Nov},
  publisher = {American Physical Society},
  doi = {10.1103/PhysRevB.86.205114},
  url = {https://link.aps.org/doi/10.1103/PhysRevB.86.205114}
}

@article{Diptiman12,
  title = {Junction between surfaces of two topological insulators},
  author = {Sen, Diptiman and Deb, Oindrila},
  journal = {Phys. Rev. B},
  volume = {85},
  issue = {24},
  pages = {245402},
  numpages = {10},
  year = {2012},
  month = {Jun},
  publisher = {American Physical Society},
  doi = {10.1103/PhysRevB.85.245402},
  url = {https://link.aps.org/doi/10.1103/PhysRevB.85.245402}
}

@article{Suri_2021,
doi = {10.1088/1361-648X/ac06ea},
url = {https://doi.org/10.1088/1361-648X/ac06ea},
year = {2021},
month = {jun},
publisher = {IOP Publishing},
volume = {33},
number = {33},
pages = {335301},
author = {Suri, Dhavala and Soori, Abhiram},
title = {Finite transverse conductance in topological insulators under an applied in-plane magnetic field},
journal = {J. Phys.: Condens. Matter},
}

@article{Milivojević_2024,
doi = {10.1088/2053-1583/ad4c73},
url = {https://doi.org/10.1088/2053-1583/ad4c73},
year = {2024},
month = {may},
publisher = {IOP Publishing},
volume = {11},
number = {3},
pages = {035025},
author = {Milivojević, Marko and Orozović, Marko and Picozzi, Silvia and Gmitra, Martin and Stavrić, Srđan},
title = {Interplay of altermagnetism and weak ferromagnetism in two-dimensional {$RuF_4$}},
journal = {2D Materials},
}

@article{Soori13,
  title = {Transport across a junction of topological insulators and a superconductor},
  author = {Soori, Abhiram and Deb, Oindrila and Sengupta, K. and Sen, Diptiman},
  journal = {Phys. Rev. B},
  volume = {87},
  issue = {24},
  pages = {245435},
  numpages = {15},
  year = {2013},
  month = {Jun},
  publisher = {American Physical Society},
  doi = {10.1103/PhysRevB.87.245435},
  url = {https://link.aps.org/doi/10.1103/PhysRevB.87.245435}
}

@article{Soori12,
  title = {Magnetic-field-induced {Fabry-P\'erot} resonances in helical edge states},
  author = {Soori, Abhiram and Das, Sourin and Rao, Sumathi},
  journal = {Phys. Rev. B},
  volume = {86},
  issue = {12},
  pages = {125312},
  numpages = {5},
  year = {2012},
  month = {Sep},
  publisher = {American Physical Society},
  doi = {10.1103/PhysRevB.86.125312},
  url = {https://link.aps.org/doi/10.1103/PhysRevB.86.125312}
}

@article{Liang2001,
author={Liang, Wenjie
and Bockrath, Marc
and Bozovic, Dolores
and Hafner, Jason H.
and Tinkham, M.
and Park, Hongkun},
title={Fabry - Perot interference in a nanotube electron waveguide},
journal={Nature},
year={2001},
month={Jun},
day={01},
volume={411},
number={6838},
pages={665-669},
issn={1476-4687},
doi={10.1038/35079517},
url={https://doi.org/10.1038/35079517}
}

@article{Cheng2011,
  title = {Topological insulators and superconductors},
  author = {Qi, Xiao-Liang and Zhang, Shou-Cheng},
  journal = {Rev. Mod. Phys.},
  volume = {83},
  issue = {4},
  pages = {1057--1110},
  numpages = {0},
  year = {2011},
  month = {Oct},
  publisher = {American Physical Society},
  doi = {10.1103/RevModPhys.83.1057},
  url = {https://link.aps.org/doi/10.1103/RevModPhys.83.1057}
}

@article{Kane10,
  title = {Colloquium: Topological insulators},
  author = {Hasan, M. Z. and Kane, C. L.},
  journal = {Rev. Mod. Phys.},
  volume = {82},
  issue = {4},
  pages = {3045--3067},
  numpages = {0},
  year = {2010},
  month = {Nov},
  publisher = {American Physical Society},
  doi = {10.1103/RevModPhys.82.3045},
  url = {https://link.aps.org/doi/10.1103/RevModPhys.82.3045}
}

@article{Sourabh14,
    author = {Barua, Sourabh and Rajeev, K. P.},
    title = {Status of surface conduction in topological insulators},
    journal = {AIP Advances},
    volume = {4},
    number = {1},
    pages = {017135},
    year = {2014},
    month = {01},
    issn = {2158-3226},
    doi = {10.1063/1.4864058},
    url = {https://doi.org/10.1063/1.4864058},
}

@article{soori2021,
doi = {10.1088/1361-648X/ac09a4},
url = {https://dx.doi.org/10.1088/1361-648X/ac09a4},
year = {2021},
month = {jun},
publisher = {IOP Publishing},
volume = {33},
number = {33},
pages = {335303},
author = {Soori, A. },
title = {Finite transverse conductance and anisotropic magnetoconductance under an applied in-plane magnetic field in two-dimensional electron gases with strong spin–orbit coupling},
journal = {J.  Phys.: Condens. Matter}
}

@article{Sahoo2023,
doi = {10.1088/1361-648X/acdbaf},
url = {https://dx.doi.org/10.1088/1361-648X/acdbaf},
year = {2023},
month = {jun},
publisher = {IOP Publishing},
volume = {35},
number = {36},
pages = {365302},
author = {Sahoo, B. K. and  Soori, A.},
title = {Transverse currents in spin transistors},
journal = {J.  Phys.: Condens. Matter}
}

@article{Archit21,
    author = {Bhardwaj, Archit and Prasad P., Syam and Raman, Karthik V. and Suri, Dhavala},
    title = {Observation of planar Hall effect in topological insulator—Bi2Te3},
    journal = {Applied Physics Letters},
    volume = {118},
    number = {24},
    pages = {241901},
    year = {2021},
    month = {06},
    issn = {0003-6951},
    doi = {10.1063/5.0053498},
    url = {https://doi.org/10.1063/5.0053498},
}

@article{Taskin2017,
author={Taskin, A. A. and  Legg, Henry F. and Yang, Fan and Sasaki, Satoshi
and Kanai, Yasushi and Matsumoto, Kazuhiko and Rosch, Achim and Ando, Yoichi},
title={Planar Hall effect from the surface of topological insulators},
journal={Nat. Commun.},
year={2017},
month={Nov},
day={07},
volume={8},
number={1},
pages={1340},
issn={2041-1723},
doi={10.1038/s41467-017-01474-8},
url={https://doi.org/10.1038/s41467-017-01474-8}
}

@article{David18,
  title = {Unconventional planar Hall effect in exchange-coupled topological insulator-ferromagnetic insulator heterostructures},
  author = {Rakhmilevich, David and Wang, Fei and Zhao, Weiwei and Chan, Moses H. W. and Moodera, Jagadeesh S. and Liu, Chaoxing and Chang, Cui-Zu},
  journal = {Phys. Rev. B},
  volume = {98},
  issue = {9},
  pages = {094404},
  numpages = {8},
  year = {2018},
  month = {Sep},
  publisher = {American Physical Society},
  doi = {10.1103/PhysRevB.98.094404},
  url = {https://link.aps.org/doi/10.1103/PhysRevB.98.094404}
}

@article{han2025,
author = {Han, Seungyun and Baek, Insu and Kim, Kyoung-Whan and Lee, Hyun-Woo and Cheon, Suik},
title = {Deterministic {Néel} Vector Switching of Altermagnets Via Magnetic Octupole Torque},
journal = {Small},
volume = {22},
number = {23},
pages = {e11790},
doi = {https://doi.org/10.1002/smll.202511790},
year = {2026}
}

@article{Sayan26,
  title = {Deterministic switching in altermagnets via asymmetric sublattice spin current},
  author = {Sarkar, Sayan and Das, Sunit and Agarwal, Amit},
  journal = {Phys. Rev. B},
  volume = {113},
  issue = {15},
  pages = {155430},
  numpages = {9},
  year = {2026},
  month = {Apr},
  publisher = {American Physical Society},
  doi = {10.1103/kzcp-wcj6},
  url = {https://link.aps.org/doi/10.1103/kzcp-wcj6}
}

@article{Zheng20,
  title = {Origin of planar Hall effect on the surface of topological insulators: Tilt of Dirac cone by an in-plane magnetic field},
  author = {Zheng, Shi-Han and Duan, Hou-Jian and Wang, Jia-Kun and Li, Jia-Yu and Deng, Ming-Xun and Wang, Rui-Qiang},
  journal = {Phys. Rev. B},
  volume = {101},
  issue = {4},
  pages = {041408(R)},
  numpages = {6},
  year = {2020},
  month = {Jan},
  publisher = {American Physical Society},
  doi = {10.1103/PhysRevB.101.041408},
  url = {https://link.aps.org/doi/10.1103/PhysRevB.101.041408}
}

@article{Tewari24,
  title = {Magnetochiral anisotropy induced nonlinear planar Hall effect in topological insulator surface states},
  author = {Marinescu, D. C. and Tewari, S.},
  journal = {Phys. Rev. B},
  volume = {109},
  issue = {20},
  pages = {205301},
  numpages = {9},
  year = {2024},
  month = {May},
  publisher = {American Physical Society},
  doi = {10.1103/PhysRevB.109.205301},
  url = {https://link.aps.org/doi/10.1103/PhysRevB.109.205301}
}

@misc{shafiei25,
      title={Planar Hall effect in ultrathin topological insulator films}, 
      author={Mohammad Shafiei and Milorad V. Milošević},
      year={2025},
      eprint={2504.10980},
      archivePrefix={arXiv},
      primaryClass={cond-mat.mes-hall},
      url={https://arxiv.org/abs/2504.10980}, 
}

@article{Rao21,
  title = {Theory for linear and nonlinear planar Hall effect in topological insulator thin films},
  author = {Rao, Wen and Zhou, Yong-Long and Wu, Yong-jia and Duan, Hou-Jian and Deng, Ming-Xun and Wang, Rui-Qiang},
  journal = {Phys. Rev. B},
  volume = {103},
  issue = {15},
  pages = {155415},
  numpages = {7},
  year = {2021},
  month = {Apr},
  publisher = {American Physical Society},
  doi = {10.1103/PhysRevB.103.155415},
  url = {https://link.aps.org/doi/10.1103/PhysRevB.103.155415}
}

@article{Smejkal22,
  title = {Beyond Conventional Ferromagnetism and Antiferromagnetism: A Phase with Nonrelativistic Spin and Crystal Rotation Symmetry},
  author = {\ifmmode \check{S}\else \v{S}\fi{}mejkal, Libor and Sinova, Jairo and Jungwirth, Tomas},
  journal = {Phys. Rev. X},
  volume = {12},
  issue = {3},
  pages = {031042},
  numpages = {16},
  year = {2022},
  month = {Sep},
  publisher = {American Physical Society},
  doi = {10.1103/PhysRevX.12.031042},
  url = {https://link.aps.org/doi/10.1103/PhysRevX.12.031042}
}

@article{smejkal22c,
  title = {Emerging Research Landscape of Altermagnetism},
  author = {\ifmmode \check{S}\else \v{S}\fi{}mejkal, L. and Sinova, J. and Jungwirth, T.},
  journal = {Phys. Rev. X},
  volume = {12},
  issue = {4},
  pages = {040501},
  numpages = {27},
  year = {2022},
  month = {Dec},
  publisher = {American Physical Society},
  doi = {10.1103/PhysRevX.12.040501},
  url = {https://link.aps.org/doi/10.1103/PhysRevX.12.040501}
}

@article{smejkal22a,
  title = {Giant and Tunneling Magnetoresistance in Unconventional Collinear Antiferromagnets with Nonrelativistic Spin-Momentum Coupling},
  author = {\ifmmode \check{S}\else \v{S}\fi{}mejkal, L. and Hellenes, A. B. and Gonz\'alez-Hern\'andez, R. and Sinova, J. and Jungwirth, T.},
  journal = {Phys. Rev. X},
  volume = {12},
  issue = {1},
  pages = {011028},
  numpages = {11},
  year = {2022},
  month = {Feb},
  publisher = {American Physical Society},
  doi = {10.1103/PhysRevX.12.011028},
  url = {https://link.aps.org/doi/10.1103/PhysRevX.12.011028}
}

@article{Cheng24,
  title = {Field-free {Josephson} diode effect in altermagnet/normal metal/altermagnet junctions},
  author = {Cheng, Qiang and Mao, Yue and Sun, Qing-Feng},
  journal = {Phys. Rev. B},
  volume = {110},
  issue = {1},
  pages = {014518},
  numpages = {12},
  year = {2024},
  month = {Jul},
  publisher = {American Physical Society},
  doi = {10.1103/PhysRevB.110.014518},
  url = {https://link.aps.org/doi/10.1103/PhysRevB.110.014518}
}

@article{Nowak25,
  title = {Magnetic-field-free parity-protected qubit based on a narrow altermagnet Josephson junction},
  author = {Vosoughi-nia, S. and Nowak, M. P.},
  journal = {Phys. Rev. B},
  volume = {114},
  issue = {22},
  pages = {225408},
  numpages = {11},
  year = {2026},
  month = {Oct},
  publisher = {American Physical Society},
  doi = {10.1103/srnr-kn2s},
  url = {https://link.aps.org/doi/10.1103/srnr-kn2s}
}

@article{sahoo26,
  title = {Giant field-free transverse Josephson diode effect in altermagnets},
  author = {Sahoo, Bijay Kumar and Soori, Abhiram},
  journal = {Phys. Rev. B},
  volume = {113},
  issue = {24},
  pages = {245433},
  numpages = {8},
  year = {2026},
  month = {Jun},
  publisher = {American Physical Society},
  doi = {10.1103/ydrg-7spp},
  url = {https://link.aps.org/doi/10.1103/ydrg-7spp}
}

\end{document}